\documentclass{aip-cp}

\usepackage[numbers]{natbib}
\usepackage{rotating}
\usepackage{graphicx}

\begin{document}

\title{The Gamma-ray Cherenkov Telescope for the Cherenkov Telescope Array}

\author[aff10,cor1]{L.~Tibaldo}
\corresp[cor1]{Corresponding author: luigi.tibaldo@mpi-hd.mpg.de}

\author[aff1]{A.~Abchiche}
\author[aff2]{D.~Allan}
\author[aff3]{J.-P.~Amans}
\author[aff2]{T.P.~Armstrong}
\author[aff4]{A.~Balzer}
\author[aff4]{D.~Berge}
\author[aff3]{C.~Boisson}
\author[aff3]{J.-J.~Bousquet}
\author[aff2]{A.M.~Brown}
\author[aff4]{M.~Bryan}
\author[aff1]{G.~Buchholtz}
\author[aff2]{P.M.~Chadwick}
\author[aff5]{H.~Costantini}
\author[aff6]{G.~Cotter}
\author[aff7]{M.K.~Daniel}
\author[aff6]{A.~De Franco}
\author[aff3]{F.~De Frondat}
\author[aff3]{J.-L.~Dournaux}
\author[aff3]{D.~Dumas}
\author[aff5]{J.-P.~Ernenwein}
\author[aff3]{G.~Fasola}
\author[aff8]{S.~Funk}
\author[aff1,aff3]{J.~Gironnet}
\author[aff2]{J.A.~Graham}
\author[aff7]{T.~Greenshaw}
\author[aff3]{O.~Hervet}
\author[aff9]{N.~Hidaka}
\author[aff10]{J.A.~Hinton}
\author[aff3]{J.-M.~Huet}
\author[aff8]{D.~Jankowsky}
\author[aff3]{I.~Jegouzo}
\author[aff8]{T.~Jogler}
\author[aff8]{M.~Kraus}
\author[aff11]{J.S.~Lapington}
\author[aff3]{P.~Laporte}
\author[aff3]{J.~Lefaucheur}
\author[aff4]{S.~Markoff}
\author[aff3]{T.~Melse}
\author[aff8]{L.~Mohrmann}
\author[aff11]{P.~Molyneux}
\author[aff2]{S.J.~Nolan}
\author[aff9]{A.~Okumura}
\author[aff11]{J.P.~Osborne}
\author[aff10]{R.D.~Parsons}
\author[aff11]{S.~Rosen}
\author[aff11]{D.~Ross}
\author[aff12]{G.~Rowell}
\author[aff13]{C.B.~Rulten}
\author[aff9]{Y.~Sato}
\author[aff3]{F.~Say\`{e}de}
\author[aff2]{J.~Schmoll}
\author[aff10]{H.~Schoorlemmer}
\author[aff3]{M.~Servillat}
\author[aff3]{H.~Sol}
\author[aff12]{V.~Stamatescu}
\author[aff4]{M.~Stephan}
\author[aff14]{R.~Stuik}
\author[aff11]{J.~Sykes}
\author[aff9]{H.~Tajima}
\author[aff11]{J.~Thornhill}
\author[aff5]{C.~Trichard}
\author[aff4]{J.~Vink}
\author[aff6]{J.J.~Watson}
\author[aff10]{R.~White}
\author[aff9]{N.~Yamane}
\author[aff3]{A.~Zech}
\author[aff8]{A.~Zink}
\author[aff10]{J.~Zorn}
\affil[aff10]{Max-Planck-Institut f\"{u}r Kernphysik, P.O. Box 103980, D 69029 Heidelberg, Germany}
\affil[aff1]{CNRS, Division technique DT-INSU, 1 Place Aristide Briand, 92190 Meudon, France}
\affil[aff2]{Department of Physics and Centre for Advanced Instrumentation, Durham University, South Road, Durham DH1 3LE, UK}
\affil[aff3]{Observatoire de Paris, CNRS, PSL University, LUTH \& GEPI, Place J. Janssen, 92195, Meudon cedex, France}
\affil[aff4]{GRAPPA, University of Amsterdam, Science Park 904, 1098 XH Amsterdam, The Netherlands}
\affil[aff5]{Aix Marseille Universit\'{e}, CNRS/IN2P3, CPPM UMR 7346 ,  163 avenue de Luminy, case 902, 13288 Marseille, France}
\affil[aff6]{Department of Physics, University of Oxford, Keble Road, Oxford OX1 3RH, UK}
\affil[aff7]{University of Liverpool, Oliver Lodge Laboratory, P.O. Box 147, Oxford Street, Liverpool L69 3BX, UK}
\affil[aff8]{Erlangen Center for Astroparticle Physics (ECAP), Erwin- Rommel-Str. 1, D 91058 Erlangen, Germany}
\affil[aff9]{Institute for Space-Earth Environmental Research, Nagoya University, Furo-cho, Chikusa-ku, Nagoya, Aichi 464-8601, Japan}
\affil[aff11]{Department of Physics and Astronomy, University of Leicester, University Road, Leicester, LE1 7RH, UK}
\affil[aff12]{School of Physical Sciences, University of Adelaide, Adelaide5005, Australia}
\affil[aff13]{Department of Physics and Astronomy, University of Minnesota, 116 Church Street, Minneapolis, MN, 55455, U.S.A.}
\affil[aff14]{Leiden Observatory, Leiden University, Postbus 9513, 2300 RA, Leiden, Netherlands}

\author{the CTA Consortium}
\eaddress[url]{http://cta-observatory.org}

\maketitle

\begin{abstract}
The Cherenkov Telescope Array (CTA) is a forthcoming ground-based observatory for very-high-energy gamma rays. CTA will consist of two arrays of imaging atmospheric Cherenkov telescopes in the Northern and Southern hemispheres, and will combine telescopes of different types to achieve unprecedented performance and energy coverage. The Gamma-ray Cherenkov Telescope (GCT) is one of the small-sized telescopes proposed for CTA to explore the energy range from a few TeV to hundreds of TeV with a field of view $\gtrsim 8^\circ$ and angular resolution of a few arcminutes. The GCT design features dual-mirror Schwarzschild-Couder optics and a compact camera based on densely-pixelated photodetectors as well as custom electronics. In this contribution we provide an overview of the GCT project with focus on prototype development and testing that is currently ongoing. We present results obtained during the first on-telescope campaign in late 2015 at the Observatoire de Paris-Meudon, during which we recorded the first Cherenkov images from atmospheric showers with the GCT multi-anode photomultiplier camera prototype. We also discuss the development of a second GCT camera prototype with silicon photomultipliers as photosensors, and plans toward a contribution to the realisation of CTA.
\end{abstract}

\section{INTRODUCTION}

The Cherenkov Telescope Array (CTA) \citep[e.g.,][]{acharya2013} is a next-generation ground-based observatory for gamma rays in the energy range from a few tens of GeV to beyond 300 TeV. The CTA concept combines a large number of imaging atmospheric Cherenkov telescopes of different sizes in two sites in the Northern and Southern hemispheres to observe the whole sky over this large energy range with unprecedented sensitivity and angular resolution.

The CTA Southern array will include 70 Small-Sized Telescopes (SSTs) \citep[e.g.,][]{montaruli2015} spread over an area of $\sim$7~km$^2$ that dominate the sensitivity in the energy range from a few TeV to 300 TeV and beyond. Multi-TeV gamma-ray showers generate a large amount of Cherenkov light, thus SSTs (with $\sim$4~m diameter reflectors) are sufficient for efficient detection. A large number  of these SSTs spread over an area of $\sim$7~km$^2$ provides the huge effective area needed to achieve the desired sensitivity at the highest energies given the rapidly declining numbers of gamma rays from sources with power-law energy spectra. With a field of view $\gtrsim 8^\circ$ and an angular resolution of a few arcminutes, the SSTs will provide the highest-resolution survey ever of the sky in this energy domain that is crucial, e.g., to understanding extreme particle accelerators in the Universe, establish the nature of the sources of Galactic cosmic rays up to the knee, and look for potential signals of new physics at the highest energies.

The Gamma-ray Cherenkov Telescope (GCT) is one of the proposed telescope designs for the CTA SST array. In this contribution we provide an overview of the GCT project, and describe the status of the prototyping effort undertaken by the GCT team. We conclude with an outlook including future plans toward the realisation of CTA.

\section{OVERVIEW OF THE GCT PROJECT}

\begin{figure}[!hbt]
  \centerline{\includegraphics[width=0.7\linewidth]{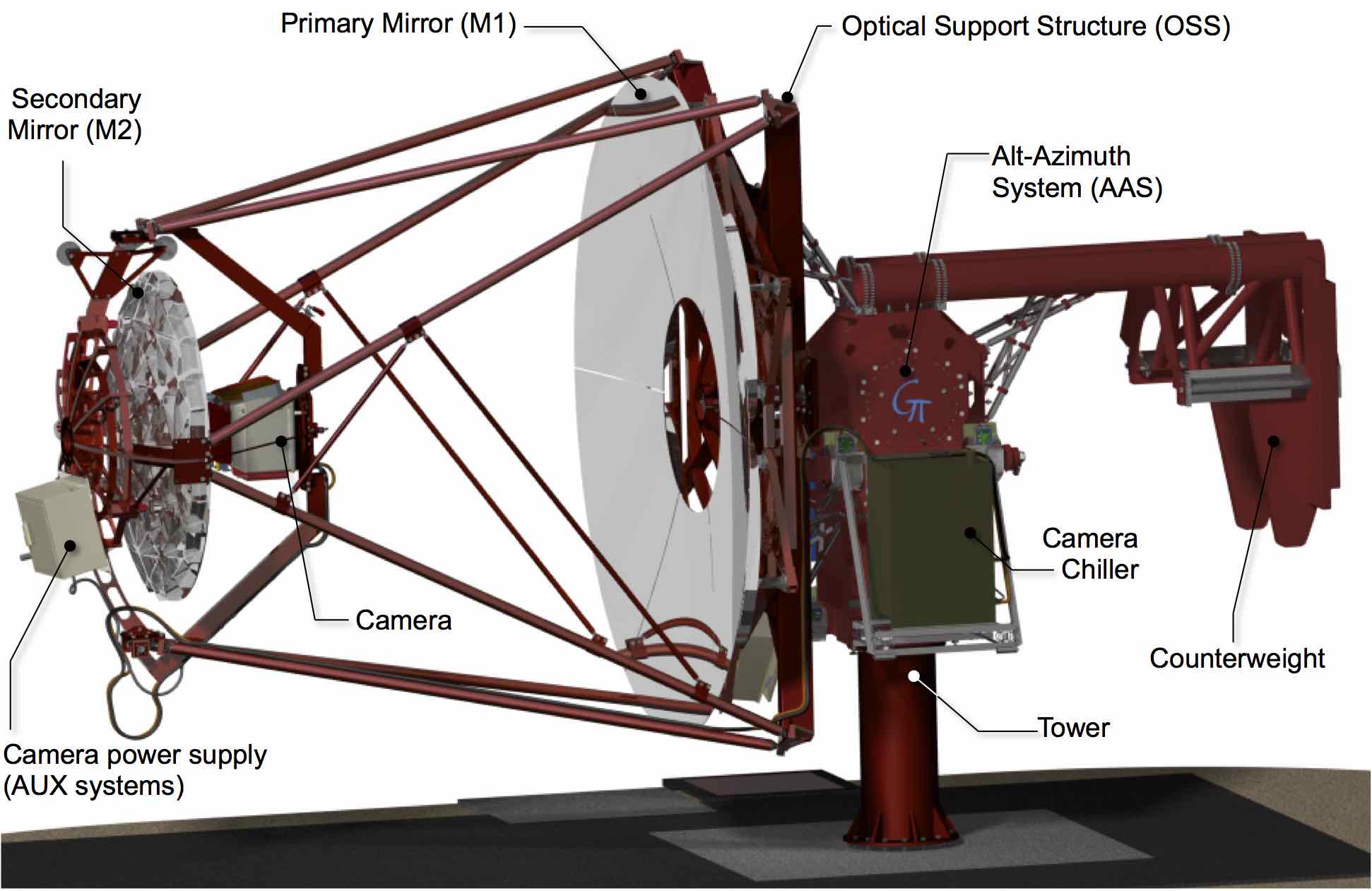}}
  \caption{A rendition of the GCT telescope.}\label{gcttel}
\end{figure}

The GCT telescope (shown in Figure~\ref{gcttel}) features a dual-mirror design with Schwarzschild-Couder optics to achieve at the same time large aperture and optimal optical performance. The primary mirror diameter is 4 m, the secondary 2 m, and the focal length is 2.3 m. The GCT camera has a photodetector surface of size $\sim$0.4~m with 2048 pixels of physical size 6-7~mm. This provides an $\sim$8$^\circ$ field of view with pixels of angular size $\sim$0.2$^\circ$. 

The GCT telescope structure consists of a foundation which supports an alt-azimuth structure controlled through drive motors. This holds the optical support structure, including support for primary and secondary dish, as well as the camera, and counterweights. For the mirrors we are considering solutions based on polished and coated aluminium or cold-slumping glass. The primary is formed of six independent petals, while the secondary is assembled as monolithic unit.

The GCT camera is based on commercially available photodetectors. The two options considered are multi-anode photomultipliers (MAPMs) and Geiger-mode avalanche photodiodes, also known as Silicon photomultipliers (SiPMs). The signal, after pre-amplification and shaping, is processed by custom electronics modules based on application-specific integrated circuits (ASICs) of the TARGET family \citep{bechtol2012target,tibaldo2015TARGET}. The TARGET ASICs sample the signal at a high and configurable rate (typically 1~Gsample~s$^{-1}$) and form trigger signals (of adjustable length) based on the analog sum of the signals in four adjacent pixels (superpixels) exceeding a configurable threshold. Analog samples, stored in a buffer with depth of 16384 cells, can be digitised and transmitted off ASIC on demand, with typical readout windows of 96 to 128 samples. Backplane electronics control power distribution and form the trigger signals for the whole camera, based on the trigger signals from the TARGET modules (typically, from the coincidence of triggers from two neighbour superpixels). The camera trigger initiates digitisation and data transmission from the TARGET modules. Data Acquisition (DACQ) boards act as a switch for transmission off camera of the data packets (in UDP format) through an optical fibre. The camera also includes LEDs for calibration purposes \citep{Brown2015flashers}, and a liquid cooling system for temperature control.

\section{THE FIRST PROTOTYPE, COMMISSIONING CAMPAIGN, AND FIRST LIGHT}

\begin{figure}[!hbt]
  \centerline{\includegraphics[width=0.75\linewidth]{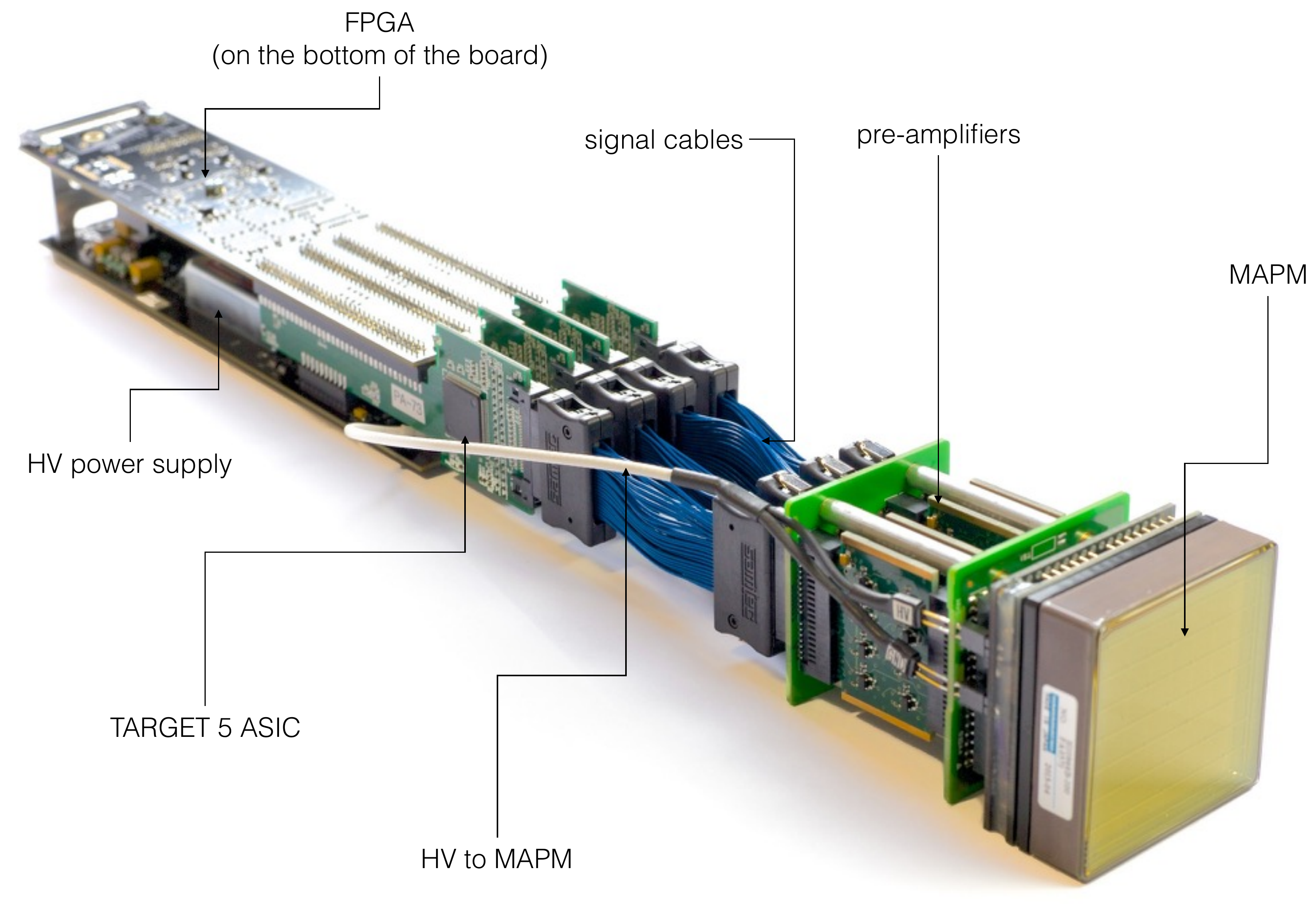}}
  \caption{One of the 32 camera modules of CHEC-M. The MAPM has 64 pixels. After pre-amplification the 64 analog signals are processed by 4 TARGET~5 ASICs (with 16 channels each). The module also produces on board the high-voltage (HV) bias for the MAPM. A single companion FPGA controls operations of the whole module and assembles data packets in UDP format to be transmitted off camera through the DACQ boards. The total module length is $\sim$40~cm.}\label{T5module}
\end{figure}

The first prototype of the GCT camera based on MAPMs, also known as CHEC-M, Compact High-Energy Camera with MAPMs, was assembled and tested in the laboratory in 2015. CHEC-M employs H10966B MAPMs from Hamamatsu and ASICs of the fifth generation of the TARGET family. Figure~\ref{T5module} shows one of the 32 camera modules that compose CHEC-M. The backplane was custom produced, while the two 1 Gbit~s$^{-1}$ DACQ boards are commercial devices from Seven Solutions. The characterisation of the camera in the laboratory included calibration of the trigger and digitiser ASICs, calibration of the photodetector-readout chain gain, and gain equalisation of the different photodetectors through tweaking of  the MAPM bias voltages.

\begin{figure}[!hbt]
  \centerline{\begin{tabular}{cc}
  \includegraphics[width=0.45\linewidth]{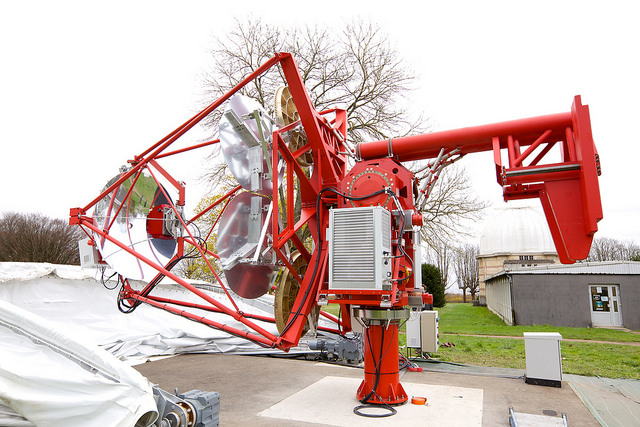}&
  \includegraphics[width=0.45\linewidth]{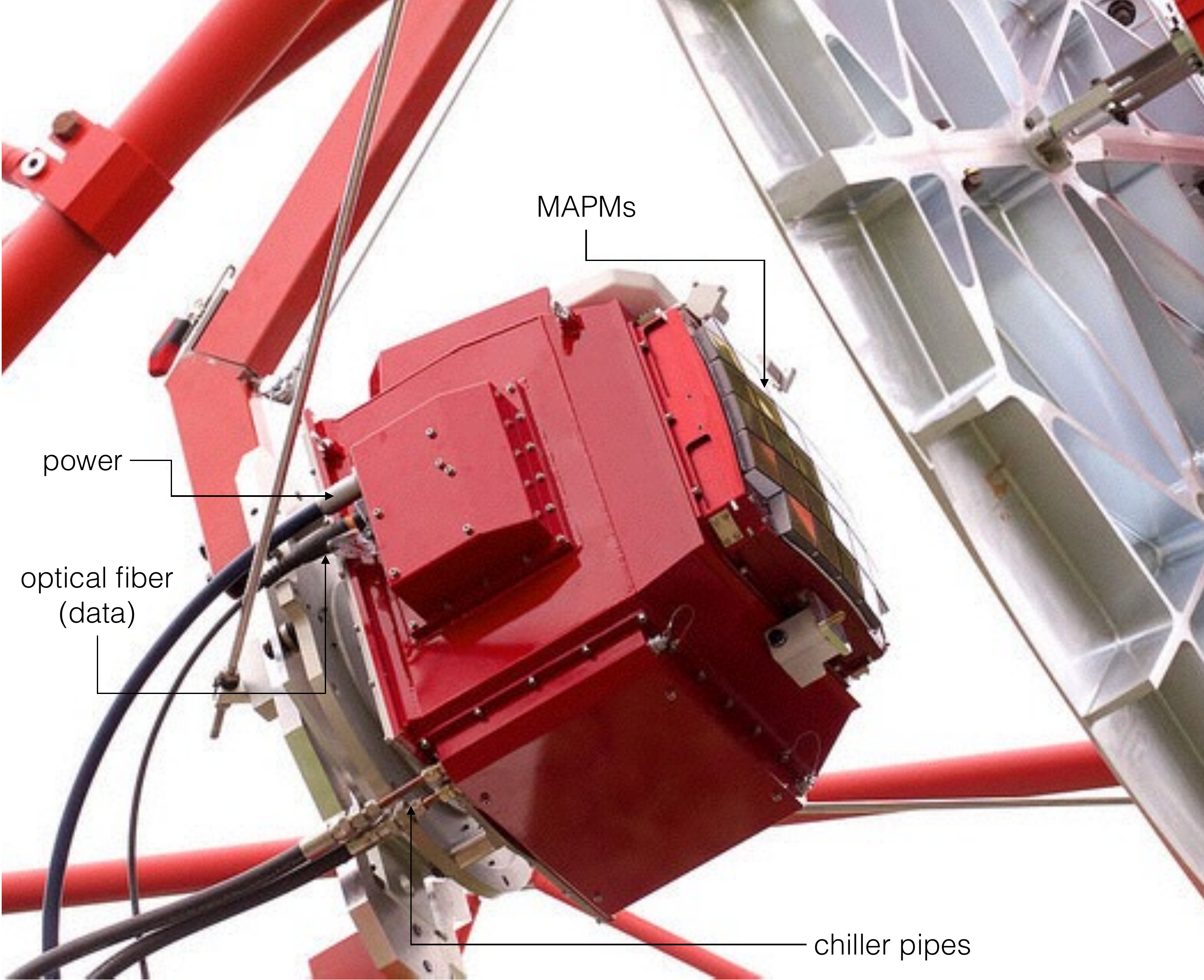}
  \end{tabular}}
  \caption{On the left, the prototype telescope at the Observatoire de Paris-Meudon with the CHEC-M camera installed during the commissioning campaign of November/December 2015 (note the circular petals used that will be replaced by trapezoidal ones in the final system). On the right, closeup of the camera.}\label{telescope}
\end{figure}

After laboratory testing, CHEC-M was installed in November 2015 on the prototype GCT telescope built at the Observatoire de Paris, in Meudon (see Figure~\ref{telescope}). The prototype telescope features only two out of six petals in the primary mirror (the other petals were replaced by dummies), and the primary petals are round rather than trapezoidal. 
After mechanical integration, the camera functionality tests were repeated on telescope, including checks on temperature stability. Data acquisition was initially tested using the calibration LEDs shining on the secondary mirror or a portable laser shining on the primary as light sources.

Owing to small delays in the system commissioning procedure, the final stage of the 2015 campaign, consisting of observations of  the night sky, took place during a few nights that were closer to full moon than anticipated, with a night-sky-background light level estimated to be 20 to 100 times brighter than at the actual CTA site. The harsh observing conditions required the use of an MAPM bias voltage of 750~V, lower than values tested earlier in the laboratory (950~V and 1100~V). This implied that the gain and trigger calibration constants were not known. Trigger thresholds were therefore set through an empirical  scan, resulting in a  trigger rate of a few~Hz.

\begin{figure}[!hbt]
  \centerline{\includegraphics[width=1\linewidth]{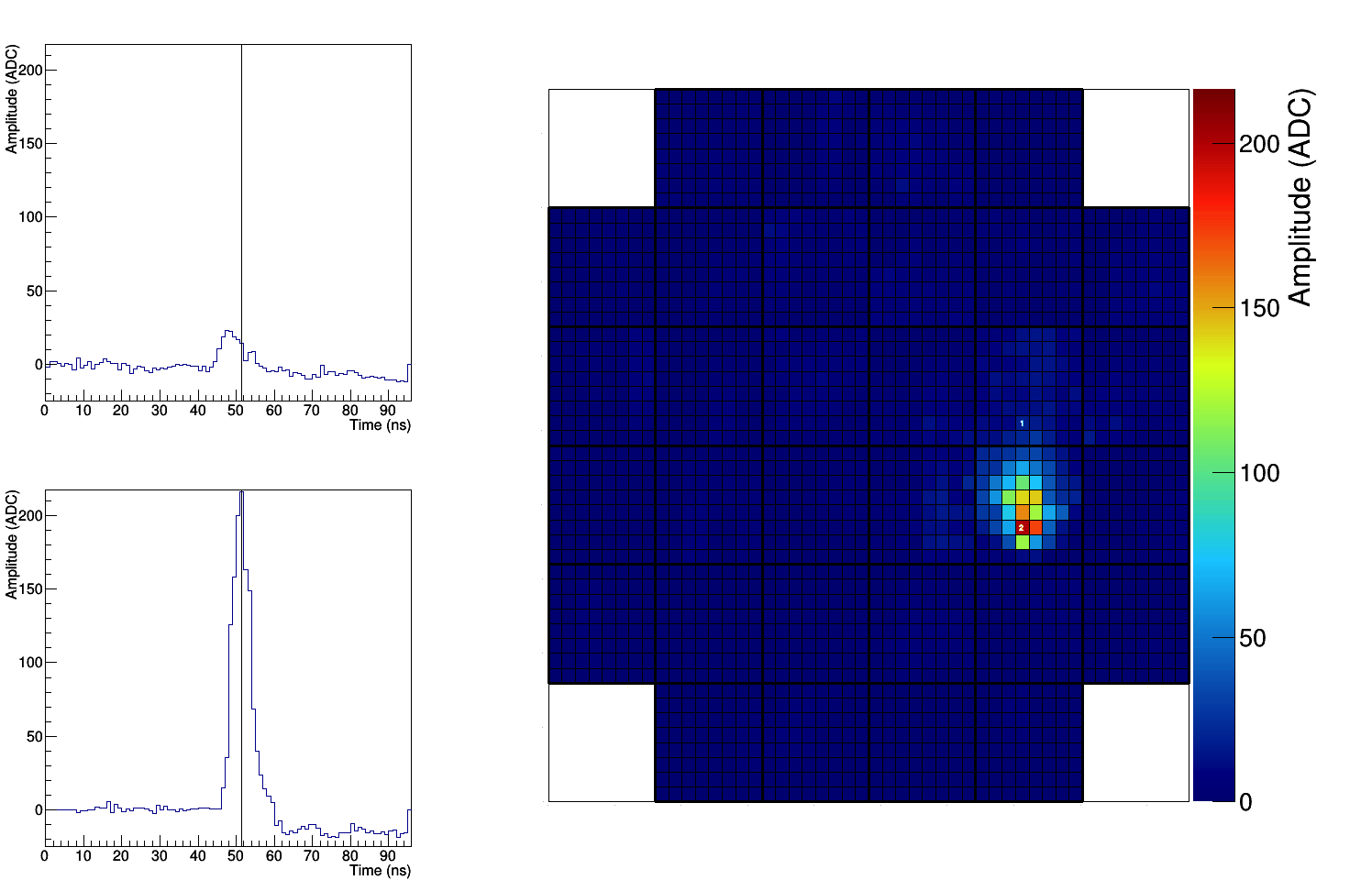}}
  \caption{A Cherenkov event recorded by CHEC-M during the on-telescope campaign of November/December 2015 at the Observatoire de Paris-Meudon. Right: a full-camera image showing the amplitude of the digital signal in each pixel (after pedestal subtraction) corresponding to the 1-ns frame when the recorded Cherenkov flash peaks. Left: waveforms from two pixels labeled as 1, top, and 2, bottom, in the right plot. The waveforms show the signals from the two pixels that were digitised and recorded over 96~ns. The vertical lines mark the time corresponding to the image on the right. Only very preliminary calibration (pedestal subtraction) was applied.}\label{Cevent}
\end{figure}

After data acquisition lasting for $\sim$10 minutes, we found a few tens of events that bore the unmistakeable appearance of Cherenkov shower images, like that shown in Figure~\ref{Cevent}.
For some of the events it was also possible to see a few~ns shift in light arrival time across the camera as expected for a Cherenkov flash from a shower inclined with respect to the telescope focal plane. The measurement of shower images by the prototype GCT telescope represents the first Cherenkov light detection using a telescope proposed for CTA, as well as the first measurement ever of Cherenkov shower images with a dual-mirror telescope. The results are discussed in more detail in these proceedings in the paper by J.~Watson et al.

\section{TOWARD THE 2016 COMMISSIONING CAMPAIGN}

After the first on-telescope campaign, CHEC-M was taken back to the laboratory to understand in more detail its performance and solve some issues that were discovered in the field.
\begin{itemize}
\item We are investigating the output data rates and the factors limiting these due to the different camera subsystems to address some issues with data packet loss; a revision of the TARGET modules and backplane electronics firmware, along with upgraded slow control and data acquisition software, solves the issue of packet loss up to a sustained data rate of 700~Hz (at the level foreseen for the GCT camera in normal operations).
\item We are characterising in more detail the triggering of the camera using a collimated light source with tuneable intensity.
\item A first prototype of a new safety board, that implements an improved safety and slow control concept (see below for a short description), will be integrated in CHEC-M for testing. 
\end{itemize}

In the mean time, the telescope prototype drive system has been fully commissioned. Optical alignment of the system was perfected, and characterisation of its optical and pointing performance is now ongoing. Preliminary results indicate that the size of the optical point-spread function on the camera focal plane is $\sim5$~mm (80\% containment diameter), which is less than the transverse size of a pixel. Additional ongoing measurements include:
\begin{itemize}
\item optical transmission of the instrument;
\item quality of the alignment of mirrors on the telescope;
\item measurement of the quality of the optical surface. 
\end{itemize}
A new campaign with CHEC-M back on the prototype telescope is anticipated for the fall of 2016. More details on the studies of the telescope prototype, and how Monte~Carlo simulations are being developed to inform the next on-telescope campaign, can be found in these proceedings in the paper by H.~Costantini et al.

\section{THE SECOND GCT CAMERA PROTOTYPE WITH SILICON PHOTOMULTIPLIERS}

We are currently developing a second camera prototype based on SiPMs, CHEC-S (Compact High-Energy Camera with SiPMs). This camera features S12642-1616PA-50 SiPMs from Hamamatsu, and a number of other upgrades and improvements:
\begin{itemize}
\item an improved mechanical design that includes focal-plane cooling to compensate for the power dissipation from the SiPMs and pre-amplifier boards, as well as a revised lid design with motorised control;
\item a new generation of TARGET ASICs, TARGET~C and T5TEA (see below for more details);
\item an improved safety and slow control concept, that includes a safety board with the capability to autonomously protect the camera subsystems from damage due to high temperature, over voltage, and high-current;
\item an upgrade to the DACQ boards to support data transmission at 10 Gbit~s$^{-1}$.
\end{itemize}

In addition to the different photodetectors, the latest-generation TARGET ASICs will provide a big step forward in performance. The original concept of a single ASIC combining sampling, digitisation, and triggering was posing a problem due to coupling between sampling and triggering operations that limited the trigger sensitivity to $\gtrsim 5$~ p.e. (with the photodetector and preamplifier gain planned for GCT). Therefore, the two functionalities were split into two separate ASICs. T5TEA provides triggering based on the same concept as TARGET~5, with a sensitivity reaching the single~p.e. level and trigger noise $\lesssim 1/4$~p.e. TARGET~C performs sampling and digitisation, with a dynamic range of 1.9~V and improved resolution with respect to TARGET 5 (cf. TARGET~7, used in the SCT telescope prototype for CTA). For more details on TARGET~C and T5TEA and the characterisation of their performance see the paper by D.~Jankowsky et al. in these proceedings. 

\section{SUMMARY AND PLANS}

We have built a full prototype of the GCT telescope proposed for CTA as well as a first camera prototype based on MAPMs. The prototype assessment and characterisation is ongoing. The full system was tested in the field at the end of 2015, achieving the first detection of Cherenkov light from atmospheric showers with an instrument proposed for CTA. A new campaign in the field is planned by the end of 2016. In the mean time, a second camera prototype is being built. This new camera uses a different type of photodetector, SiPMs, and features many improvements to the camera mechanics and electronics. It will be assembled in early 2017, and tested in the laboratory and subsequently on the prototype telescope. Thereby, we will verify that GCT meets all the requirements for CTA.

CTA observatory construction is planned to start in 2017. The first phase, dubbed pre-production, will consist of the deployment of $\sim$10\% of the total number of each type of telescope on the final site for final testing and characterisation before the mass production phase. We propose to deploy 3~GCT telescopes on site in 2018 as part of the pre-production phase for CTA. Later, we propose to contribute up to 35 SST to the CTA array.

\section{ACKNOWLEDGMENTS}
We gratefully acknowledge support from the agencies and organisations 
listed under Funding Agencies at this website: http://www.cta-observatory.org/.


\bibliographystyle{aipnum-cp}%

\begin{thebibliography}{4}%
\makeatletter
\providecommand \@ifxundefined [1]{%
 \@ifx{#1\undefined}
}%
\providecommand \@ifnum [1]{%
 \ifnum #1\expandafter \@firstoftwo
 \else \expandafter \@secondoftwo
 \fi
}%
\providecommand \@ifx [1]{%
 \ifx #1\expandafter \@firstoftwo
 \else \expandafter \@secondoftwo
 \fi
}%
\providecommand \natexlab [1]{#1}%
\providecommand \enquote  [1]{``#1''}%
\providecommand \bibnamefont  [1]{#1}%
\providecommand \bibfnamefont [1]{#1}%
\providecommand \citenamefont [1]{#1}%
\providecommand \href@noop [0]{\@secondoftwo}%
\providecommand \href [0]{\begingroup \@sanitize@url \@href}%
\providecommand \@href[1]{\@@startlink{#1}\@@href}%
\providecommand \@@href[1]{\endgroup#1\@@endlink}%
\providecommand \@sanitize@url [0]{\catcode `\$12\catcode `\&12\catcode
  `\#12\catcode `\^12\catcode `\_12\catcode `\%12\relax}%
\providecommand \@@startlink[1]{}%
\providecommand \@@endlink[0]{}%
\providecommand \url  [0]{\begingroup\@sanitize@url \@url }%
\providecommand \@url [1]{\endgroup\@href {#1}{\urlprefix }}%
\providecommand \urlprefix  [0]{URL }%
\providecommand \Eprint [0]{\href }%
\providecommand \doibase [0]{http://dx.doi.org/}%
\providecommand \selectlanguage [0]{\@gobble}%
\providecommand \bibinfo  [0]{\@secondoftwo}%
\providecommand \bibfield  [0]{\@secondoftwo}%
\providecommand \translation [1]{[#1]}%
\providecommand \BibitemOpen [0]{}%
\providecommand \bibitemStop [0]{}%
\providecommand \bibitemNoStop [0]{.\EOS\space}%
\providecommand \EOS [0]{\spacefactor3000\relax}%
\providecommand \BibitemShut  [1]{\csname bibitem#1\endcsname}%
\let\auto@bib@innerbib\@empty
\bibitem [{\citenamefont {{Acharya}}\ \emph {et~al.}(2013)\citenamefont
  {{Acharya}}, \citenamefont {{Actis}}, \citenamefont {{Aghajani}},
  \citenamefont {{Agnetta}}, \citenamefont {{Aguilar}}, \citenamefont
  {{Aharonian}}, \citenamefont {{Ajello}}, \citenamefont {{Akhperjanian}},
  \citenamefont {{Alcubierre}}, \citenamefont {{Aleksi{\'c}}},\ and\
  \citenamefont {et~al.}}]{acharya2013}%
  \BibitemOpen
  \bibfield  {author} {\bibinfo {author} {\bibfnamefont {B.~S.}\ \bibnamefont
  {{Acharya}}}, \bibinfo {author} {\bibfnamefont {M.}~\bibnamefont {{Actis}}},
  \bibinfo {author} {\bibfnamefont {T.}~\bibnamefont {{Aghajani}}}, \bibinfo {author}
  {\bibnamefont {et~al.}},\ }\href {\doibase
  10.1016/j.astropartphys.2013.01.007} {\bibfield  {journal} {\bibinfo
  {journal} {Astroparticle Physics}\ }\textbf {\bibinfo {volume} {43}},\
  \unskip\ \bibinfo {pages} {3--18} March (\bibinfo {year} {2013})}\BibitemShut
  {NoStop}%
\bibitem [{\citenamefont {{Montaruli}}, \citenamefont {{Pareschi}},\ and\
  \citenamefont {{Greenshaw}}(2015)}]{montaruli2015}%
  \BibitemOpen
  \bibfield  {author} {\bibinfo {author} {\bibfnamefont {T.}~\bibnamefont
  {{Montaruli}}}, \bibinfo {author} {\bibfnamefont {G.}~\bibnamefont
  {{Pareschi}}}, \ and\ \bibinfo {author} {\bibfnamefont {T.}~\bibnamefont
  {{Greenshaw}}},\ }\href@noop {} {\bibfield  {journal} {\bibinfo  {journal}
  {PoS(ICRC2015)1043}\ August} (\bibinfo {year} {2015})},\ \Eprint
  {http://arxiv.org/abs/1508.06472} {arXiv:1508.06472}\BibitemShut {NoStop}%
\bibitem [{\citenamefont {{Bechtol}}\ \emph {et~al.}(2012)\citenamefont
  {{Bechtol}}, \citenamefont {{Funk}}, \citenamefont {{Okumura}}, \citenamefont
  {{Ruckman}}, \citenamefont {{Simons}}, \citenamefont {{Tajima}},
  \citenamefont {{Vandenbroucke}},\ and\ \citenamefont
  {{Varner}}}]{bechtol2012target}%
  \BibitemOpen
  \bibfield  {author} {\bibinfo {author} {\bibfnamefont {K.}~\bibnamefont
  {{Bechtol}}}, \bibinfo {author} {\bibfnamefont {S.}~\bibnamefont {{Funk}}},
  \bibinfo {author} {\bibfnamefont {A.}~\bibnamefont {{Okumura}}}, \bibinfo
  {author} {\bibfnamefont {L.~L.}\ \bibnamefont {{Ruckman}}}, \bibinfo {author}
  {\bibfnamefont {A.}~\bibnamefont {{Simons}}}, \bibinfo {author}
  {\bibfnamefont {H.}~\bibnamefont {{Tajima}}}, \bibinfo {author}
  {\bibfnamefont {J.}~\bibnamefont {{Vandenbroucke}}}, \ and\ \bibinfo {author}
  {\bibfnamefont {G.~S.}\ \bibnamefont {{Varner}}},\ }\href {\doibase
  10.1016/j.astropartphys.2012.05.016} {\bibfield  {journal} {\bibinfo
  {journal} {Astroparticle Physics}\ }\textbf {\bibinfo {volume} {36}},\
  \unskip\ \bibinfo {pages} {156--165} August (\bibinfo {year} {2012})},\
  \Eprint {http://arxiv.org/abs/1105.1832} {arXiv:1105.1832}\BibitemShut {NoStop}%
\bibitem [{\citenamefont {{Tibaldo}}\ \emph {et~al.}(2015)\citenamefont
  {{Tibaldo}}, \citenamefont {{Vandenbroucke}}, \citenamefont {{Albert}},
  \citenamefont {{Funk}}, \citenamefont {{Kawashima}}, \citenamefont {{Kraus}},
  \citenamefont {{Okumura}}, \citenamefont {{Sapozhnikov}}, \citenamefont
  {{Tajima}}, \citenamefont {{Varner}}, \citenamefont {{Wu}}, \citenamefont
  {{Zink}},\ and\ \citenamefont {{CTA consortium}}}]{tibaldo2015TARGET}%
  \BibitemOpen
  \bibfield  {author} {\bibinfo {author} {\bibfnamefont {L.}~\bibnamefont
  {{Tibaldo}}}, \bibinfo {author} {\bibfnamefont {J.~A.}\ \bibnamefont
  {{Vandenbroucke}}}, \bibinfo {author} {\bibfnamefont {A.~M.}\ \bibnamefont
  {{Albert}}}, \bibinfo {author} {\bibfnamefont {S.}~\bibnamefont {{Funk}}},
  \bibinfo {author} {\bibfnamefont {T.}~\bibnamefont {{Kawashima}}}, \bibinfo
  {author} {\bibfnamefont {M.}~\bibnamefont {{Kraus}}}, \bibinfo {author}
  {\bibfnamefont {A.}~\bibnamefont {{Okumura}}}, \bibinfo {author}
  {\bibfnamefont {L.}~\bibnamefont {{Sapozhnikov}}}, \bibinfo {author}
  {\bibfnamefont {H.}~\bibnamefont {{Tajima}}}, \bibinfo {author}
  {\bibfnamefont {G.~S.}\ \bibnamefont {{Varner}}}, \bibinfo {author}
  {\bibfnamefont {T.}~\bibnamefont {{Wu}}}, \bibinfo {author} {\bibfnamefont
  {A.}~\bibnamefont {{Zink}}}, }\href@noop {} {\bibfield  {journal}
  {\bibinfo  {journal} {PoS(ICRC2015)932}\ August} (\bibinfo {year} {2015})},\
 {arXiv:1508.06296}\BibitemShut {NoStop}%
 \bibitem{Brown2015flashers}%
  \BibitemOpen
  \bibfield  {author} {\bibinfo {author} {\bibfnamefont {A.~M.}~\bibnamefont
  {{Brown}}}, \bibinfo {author} {\bibfnamefont {T}~\bibnamefont
  {{Armstrong}}}, \bibinfo {author} {\bibfnamefont {P.~M.}~\bibnamefont
  {{Chadwick}}}, \bibinfo {author} {\bibfnamefont {M.}~\bibnamefont
  {{Daniel}}}, \bibinfo {author} {\bibfnamefont {R.}~\bibnamefont
  {{White}}}, }\href@noop {} {\bibfield  {journal}
  {\bibinfo  {journal} {PoS(ICRC2015)934}\ August} (\bibinfo {year} {2015})}\BibitemShut {NoStop}%
\end{thebibliography}
%

\end{document}